\documentclass[twocolumn,showpacs,preprintnumbers,amsmath,amssymb]{revtex4}
\usepackage{graphicx}% Include figure files
\usepackage{dcolumn}% Align table columns on decimal point
\usepackage{bm}% bold math

\newcommand{\bef}{\begin{figure}}
\newcommand{\eef}{\end{figure}}

\newcommand{\be}{\begin{equation}}
\newcommand{\ee}{\end{equation}}
\newcommand{\bea}{\begin{eqnarray}}
\newcommand{\eea}{\end{eqnarray}}
\begin{document}

\title{Energy dependence of elliptic flow from heavy-ion collision models}

\author{Md. Nasim$^1$, Lokesh Kumar$^2$, Pawan Kumar Netrakanti$^3$, and Bedangadas Mohanty$^1$ }
\affiliation{$^1$Variable Energy Cyclotron Centre, Kolkata 700064, India, $^2$Kent State University, Kent, Ohio 44242, USA, and $^3$Bhabha Atomic Research Centre, Mumbai 400 085, India}

\date{\today}
\begin{abstract}
We have compared the experimental data on charged particle elliptic flow 
parameter ($v_{2}$) in Au+Au collisions at midrapidity for 
$\sqrt{s_{\mathrm {NN}}}$ = 9.2, 19.6, 62.4 and 200 GeV with results 
from various models in heavy-ion collisions like UrQMD, AMPT, and HIJING. 
We observe that the average $v_{2}$ ($\langle v_{2} \rangle$) from the 
transport model UrQMD agrees well with the measurements at 
$\sqrt{s_{\mathrm {NN}}}$ = 9.2 GeV but increasingly falls short 
of the experimental $\langle v_{2} \rangle$ values as the beam energy 
increases. The difference in $\langle v_{2} \rangle$ being of the order 
of 60\% at $\sqrt{s_{\mathrm {NN}}}$ = 200 GeV. The $\langle v_{2} \rangle$ 
results from HIJING is consistent with zero, while those from AMPT with 
default settings, a model based on HIJING with additional initial and final 
state rescattering effects included, gives a $\langle v_{2} \rangle$ 
value of about 4\% for all the beam energies studied. This is in contrast to
increase in $\langle v_{2} \rangle$ with beam energy for the experimental data.
A different version of the AMPT model, which includes partonic effects and quark 
coalescence as a mechanism of hadronization, gives higher values of 
$\langle v_{2} \rangle$ among the models studied and is in agreement with 
the measured $\langle v_{2} \rangle$ values at  $\sqrt{s_{\mathrm {NN}}}$ = 200 GeV.  
These studies show that the experimental $\langle v_{2} \rangle$  has substantial 
contribution from partonic interactions at $\sqrt{s_{\mathrm {NN}}}$ = 200 GeV 
whose magnitude reduces with decrease in beam energy. We also compare the available
data on the transverse momentum and pseudorapidity dependence of $v_{2}$
to those from the above models.
\end{abstract}
\maketitle

Elliptic flow ($v_{2}$) measured in heavy-ion collisions are believed to 
arise because of the pressure gradient developed when two nuclei collides
at non-zero impact parameters followed by subsequent interactions among
the constituents~\cite{flow1}. Within a hydrodynamical framework, the $v_{2}$ has been
shown to be sensitive to the equation of state of the system formed in the 
collisions~\cite{flow2}. Recent data on $v_{2}$ for baryons and mesons
when measured as a function of $m_{\mathrm T} - m_{h}$, where 
$m_{\mathrm T}$ (= $\sqrt{p_{\mathrm T}^2 + m_{h}^{2}}$) is the transverse mass,
$p_{\mathrm T}$ is the transverse momentum and $m_{h}$ is the mass of the hadron,
show a unique scaling at $\sqrt{s_{\mathrm {NN}}}$ = 200 GeV.
When $v_{2}$ and $m_{\mathrm T} - m_{h}$ are scaled by the number of 
constituent quarks for a hadron, the $v_{2}$ values follow an universal 
scaling for all the measured hadrons~\cite{rhicflow}. This observation,
refered to as the number of constituent quark scaling, has been interpreted 
as the collectivity being developed at the 
partonic stage of the evolution of the system in heavy-ion collisions~\cite{voloshin,starphiflow}. 
Further it has been shown that the pseudorapidity ($\eta$) dependence of $v_{2}$
for charged hadrons shows a longitudinal scaling as observed for the 
multiplicity distributions in these collisions~\cite{phobosflow,pmdmult}. 

The elliptic flow parameter is defined as the $2^{\mathrm {nd}}$ Fourier coefficient $v_{2}$ 
of the particle distributions in emission azimuthal angle ($\phi$) with respect to
the reaction plane angle ($\Psi$)~\cite{voloshinzhang}, and can be written as
\begin{equation} \frac{dN}{d\phi} \propto
%1+2\sum v_2\cos(2(\phi - \Psi)), \end{equation}
1+2 v_2\cos(2(\phi - \Psi)). \end{equation}
For a given rapidity window the second coefficient is
\begin{equation}
v_{2}=\langle\cos(2(\phi-\Psi))\rangle=\langle\frac{p_x^2-p_y^2}{p_x^2+p_y^2}\rangle,
\end{equation}
where $p_x$ and $p_y$ are the $x$ and $y$ component of the particle momenta.
%%%%%%%%%%%%%% Fig. 2 %%%%%%%%%%%%%%%%%%%%%%%%%%%%%
\bef
\begin{center}
\includegraphics[scale=0.4]{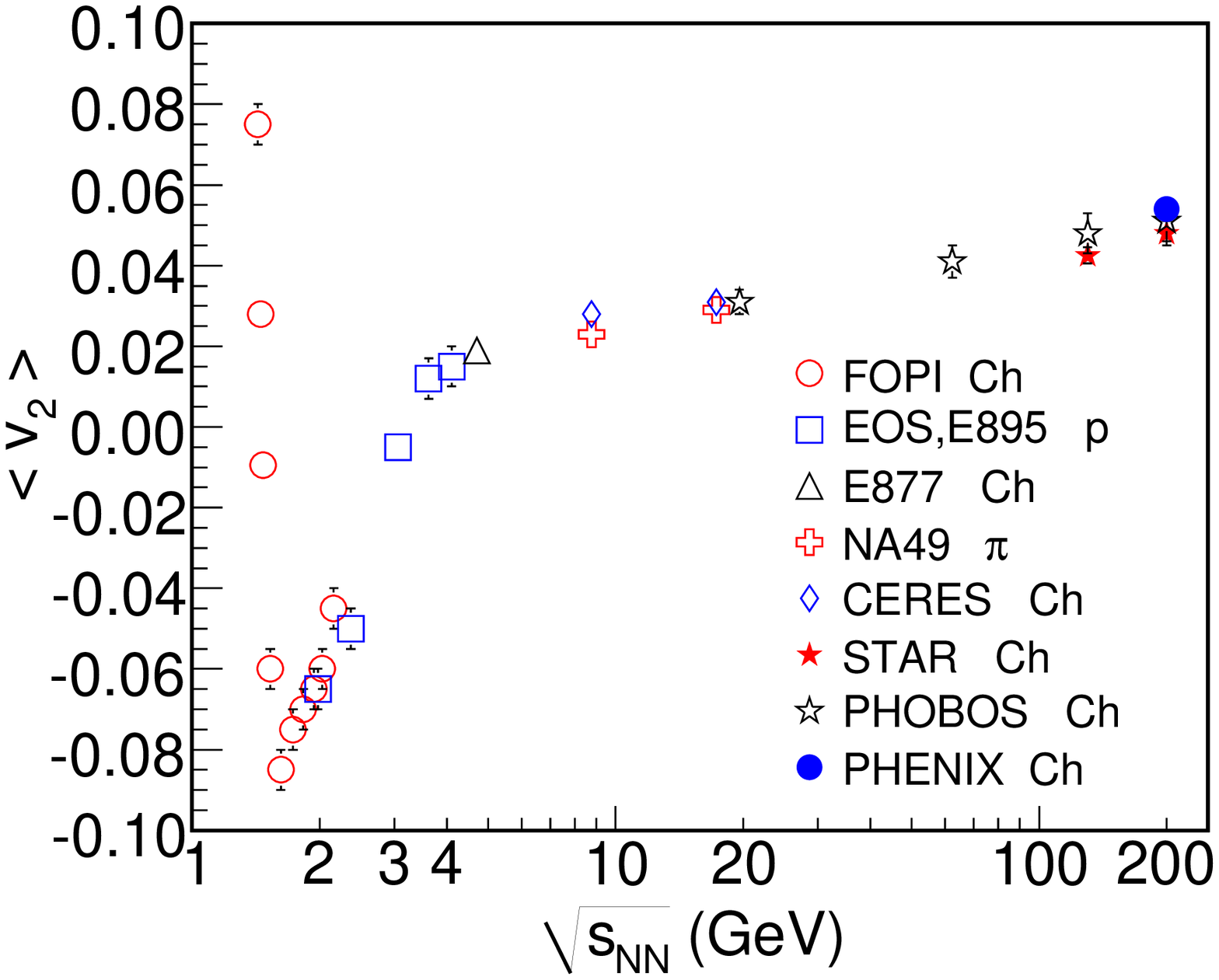}
\caption{(Color online) Average elliptic flow ($\langle v_{2} \rangle$) as a function
of beam energy. The results are shown for charged particles from RHIC experiments of 
STAR~\cite{starflow}, PHENIX~\cite{phenixflow} and PHOBOS~\cite{phobosflow}, 
SPS experiments of CERES~\cite{ceres} and NA49 (charged pions)~\cite{na49}, AGS 
experiments of E877~\cite{e877} and E895~\cite{e895} (proton) and FOPI~\cite{fopi}.}
\label{fig1}
\end{center}
\eef
%%%%%%%%%% End of Fig.1 %%%%%%%%%%%%%%%%%%%%%%%%%%%%%

The Relativistic Heavy Ion Collider (RHIC) has undertaken a beam energy scan program
to look for change in observations of various measurements as a function of
beam energy to establish the partonic phase at higher energy collisions~\cite{9gev}.
In such a program, the energy dependence of elliptic flow will play a crucial
role. The compilation of available average charged particle $\langle v_{2} \rangle$ as a
function of beam energy~\cite{art} is shown in Fig.~\ref{fig1}. 
A non-monotonic dependence
of  $\langle v_{2} \rangle$ vs. $\sqrt{s_{\mathrm {NN}}}$ is observed.

At lower energies the negative $\langle v_{2} \rangle$ is attributed to out-of-plane 
squeeze-out phenomena~\cite{art,gut}. In this case the elliptical shape of the particle 
transverse momentum distribution at midrapidity is elongated in the direction perpendicular 
to the reaction plane and interpreted as due to shadowing by spectator nucleons.
At high energies, the longitudinal size of the Lorentz contracted nuclei becomes 
negligible compared to its transverse size. This decreases the crossing time scales of the 
two nuclei. The shadowing effect goes away and elliptic flow fully develops in-plane, 
leading to a positive value of $\langle v_{2} \rangle$. In this work, we concentrate
on understanding this positive $\langle v_{2} \rangle$ values at high energies measured
at RHIC by comparing to available models of heavy-ion collisions.

Various observables are compared to theoretical calculations to understand
the physical mechanism behind the measurements. Some of the frequently
used models in heavy-ion collisions are UrQMD~\cite{urqmd}, AMPT~\cite{ampt} 
and HIJING~\cite{hijing}. 
HIJING (or Heavy Ion Jet Interaction Generator) is an event generator
for heavy-ion collisions. It is a perturbative QCD inspired model which
produces multiple minijet partons, these later get transformed into string 
configurations and then fragment to hadrons. The fragmentation is 
based on the Lund jet fragmentation model~\cite{lund}.
A parameterized parton distribution function inside a nucleus is used to take 
into account parton shadowing. Such a model does not have the mechanism to generate
elliptic flow, however it would be interesting to know how much of the 
correlations among hadrons from the minijets  contribute to the $\langle v_{2} \rangle$. 

The AMPT (A Multi Phase Transport model) uses the same initial conditions as in HIJING.
However the minijet partons are made to undergo scattering before they are allowed
to fragment into hadrons. The string melting (SM) version of the AMPT model 
(labeled here as AMPT-SM) is based on the idea that for energy densities beyond 
a critical value of $\sim$ 1 GeV/$\rm {fm}^{3}$,
it is difficult to visualize the coexistence of strings (or hadrons) and partons. 
Hence the need to melt the strings to partons. This is done by converting the mesons to
a quark and anti-quark pair, baryons to three quarks etc.  The scattering of the quarks
are based on parton cascade~\cite{ampt}. Once the interactions stop, the partons then 
hadronizes through the mechanism of parton coalescence. The interactions between
the minijet partons in AMPT model and those between partons in the AMPT-SM
model could give rise to substantial $\langle v_{2} \rangle$. Agreement between
the data and the results from AMPT-SM would indicate the contribution of partonic 
interactions to the observed $\langle v_{2} \rangle$. 

The UrQMD (Ultra relativistic Quantum 
Molecular Dynamics) model is based on a microscopic transport theory where the
phase space description of the reactions are important. It allows for the propagation
of all hadrons on classical trajectories in combination with stochastic binary 
scattering, color string formation and resonance decay. It incorporates baryon-baryon,
meson-baryon and meson-meson interactions, the collisional term includes more than 50 
baryon species and 45 meson species. The comparison of the data on $\langle v_{2} \rangle$ 
to those obtained from UrQMD model will tell about the contribution from the hadronic phase.

In this paper we compare the $\langle v_{2} \rangle$ for charged particles 
at midrapidity for $\sqrt{s_{\mathrm {NN}}}$ = 9.2, 19.6, 62.4 and 200 GeV with those from UrQMD
(ver.2.3 ), AMPT (ver. 1.11) and HIJING (ver. 1.35) models. The parton-parton 
interaction cross section in the string melting version of the AMPT is taken to be 10 mb.  
We also discuss the $p_{\mathrm T}$ and $\eta$ 
dependences of $\langle v_{2} \rangle$.

%%%%%%%%%%%%%% Fig. 2 %%%%%%%%%%%%%%%%%%%%%%%%%%%%%
\bef
\begin{center}
\includegraphics[scale=0.4]{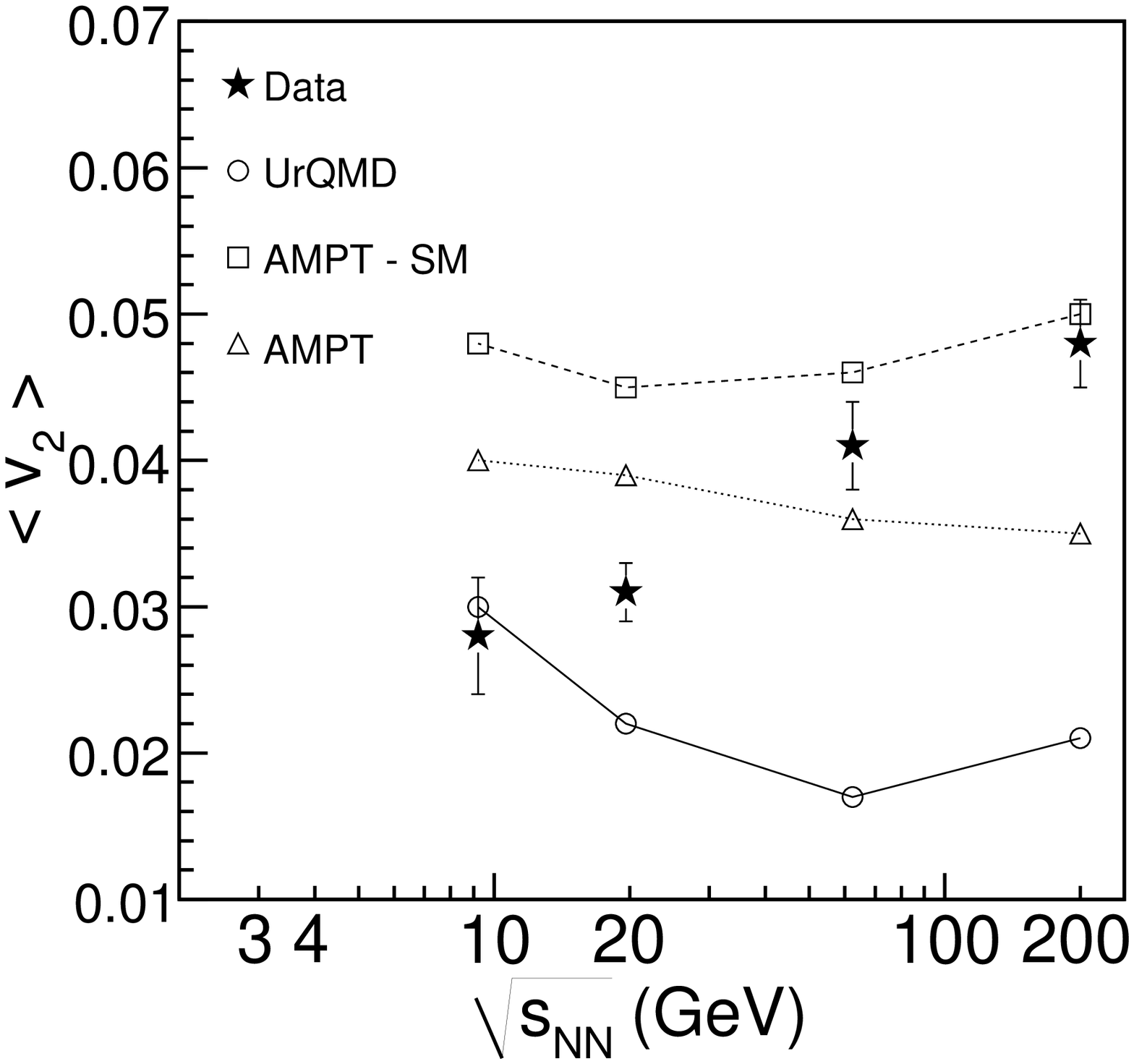}
\caption{$\langle v_{2} \rangle$ for charged particles at midrapidity 
for minimum bias collisions at $\sqrt{s_{\mathrm {NN}}}$ = 9.2, 19.6, 62.4 and 200 GeV~\cite{9gev,starflow} 
compared to corresponding results from AMPT and UrQMD model calculations.}
\label{fig2}
\end{center}
\eef
%%%%%%%%%% End of Fig.2 %%%%%%%%%%%%%%%%%%%%%%%%%%%%%
Figure~\ref{fig2} shows the $\langle v_{2} \rangle$ for charged particles 
at midrapidity for various $\sqrt{s_{\mathrm {NN}}}$ for minimum bias 
(0-80\%) collisions~\cite{starflow}. The results for $\sqrt{s_{\mathrm {NN}}}$ = 9.2 GeV
are for minimum bias 0-60\% collisions~\cite{9gev}. The $\langle v_{2} \rangle$ value increases 
linearly from about 3\% at 9.2 GeV to about 5\% at 200 GeV. 
The experimental data are compared to $\langle v_{2} \rangle$
calculated from UrQMD, AMPT and AMPT-SM with default settings.
The centrality selection is same for data and the models.
In contrast to observations from the data, the model $\langle v_{2} \rangle$ 
values either remain constant or decreases slightly with increasing $\sqrt{s_{\mathrm {NN}}}$.
The $\langle v_{2} \rangle$ value from UrQMD at 9.2 GeV and those
from AMPT-SM at 200 GeV are in good agreement with the data.
The $\langle v_{2} \rangle$ values from AMPT lie intermediate 
to those from UrQMD and AMPT-SM. If we assume the $\langle v_{2} \rangle$
values from UrQMD to be the contribution from hadronic phase, then
this contribution ($v_{2}^{UrQMD}$/$v_{2}^{data}$) 
varies from 100\% to about 40\% of the measured
$\langle v_{2} \rangle$ as beam energy increases from 9.2 GeV to 200 GeV. 
The higher values of $\langle v_{2} \rangle$ in data
indicate the possible contribution that can come in such transport
models due to inclusion of initial/final state scattering effects and/or due to
partonic interactions. Comparison with AMPT-SM reflects that at 62.4 
and 200 GeV, the $\langle v_{2} \rangle$ has contributions from partonic
interactions. We have estimated the $\langle v_{2} \rangle$ values from
HIJING and found them to be consistent with zero. The $\langle v_{2} \rangle$
values obtained from HIJING are  -0.00006 $\pm$ 0.0004, 0.0001 $\pm$ 0.0005 
and -0.002 $\pm$ 0.0005 at 200, 62.4 and 19.6 GeV respectively.

We now present the differential $v_{2}$ by comparing the $p_{\mathrm T}$
and $\eta$ dependence from data with those from the models. 
%%%%%%%%%%%%%% Fig. 3 %%%%%%%%%%%%%%%%%%%%%%%%%%%%%
\bef
\begin{center}
\includegraphics[scale=0.4]{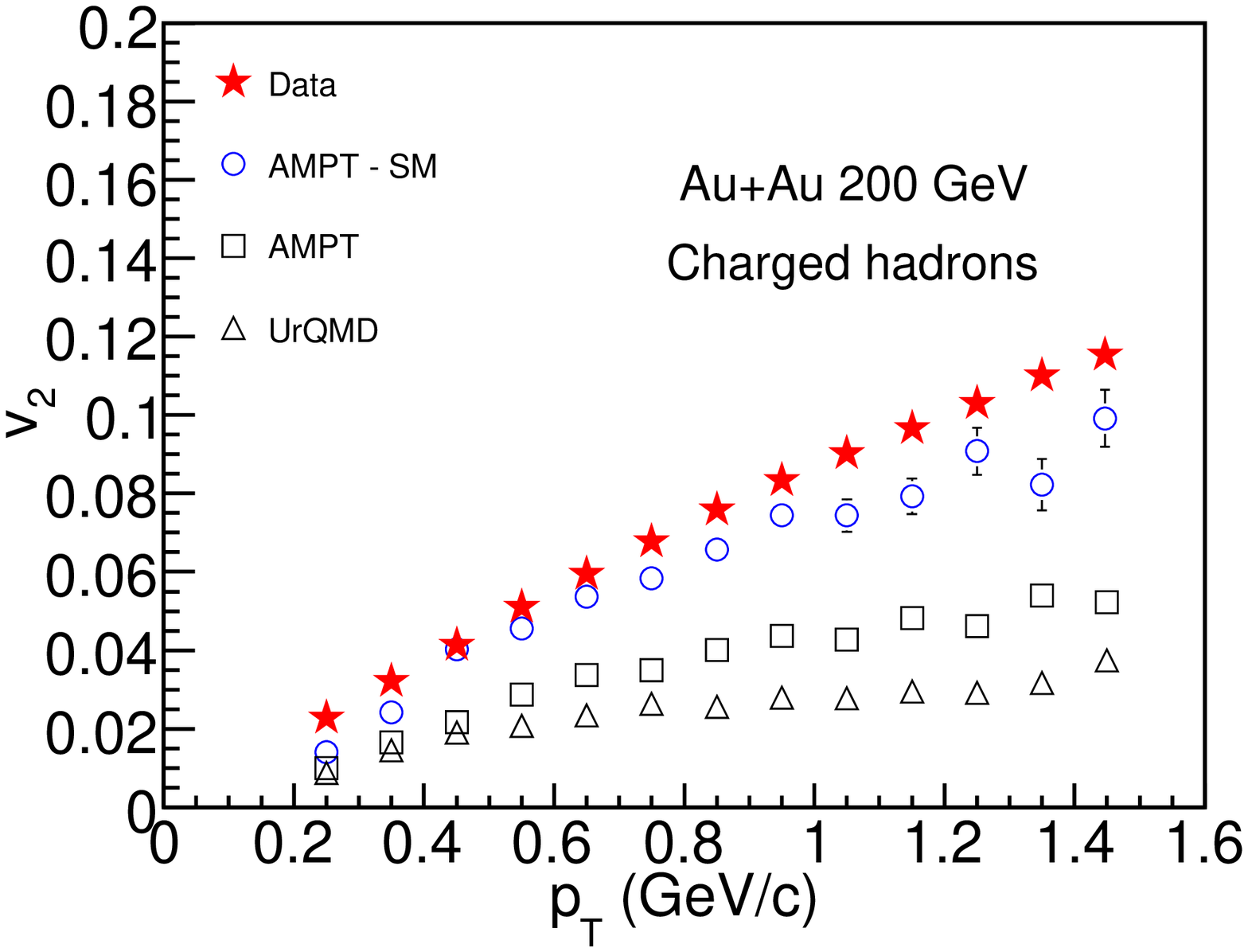}
\includegraphics[scale=0.4]{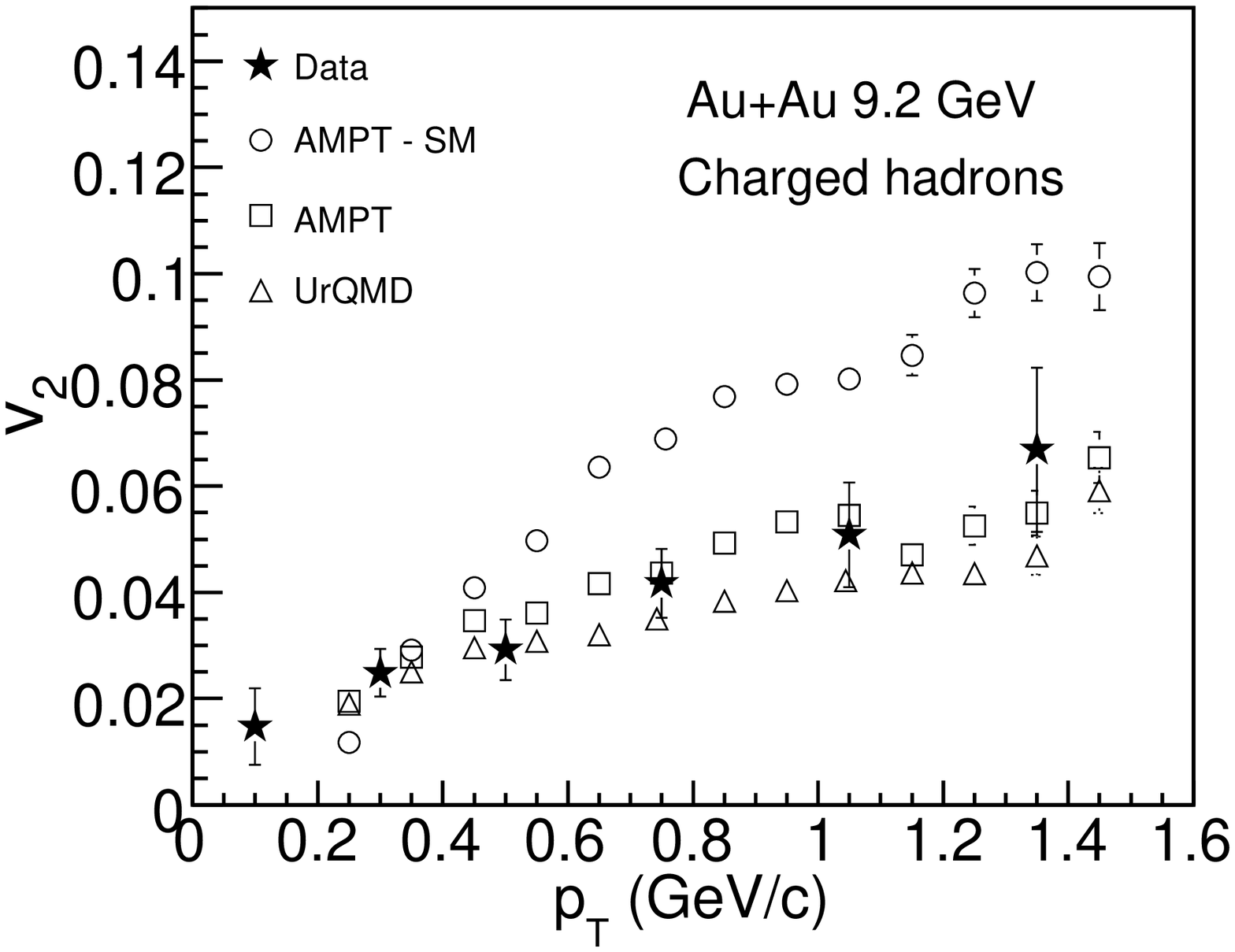}
\caption{(Color online) $v_{2}$ as a function of $p_{\mathrm T}$ at 
midrapidity for Au+Au collisions at $\sqrt{s_{\mathrm {NN}}}$ = 200~\cite{starflow} 
and 9.2 GeV~\cite{9gev}.
The experimental results are compared to corresponding $v_{2}$ values from 
UrQMD and AMPT models. The errors on the data points are statistical.}
\label{fig3}
\end{center}
\eef
%%%%%%%%%% End of Fig.3 %%%%%%%%%%%%%%%%%%%%%%%%%%%%%
%%%%%%%%%%%%%% Fig. 4 %%%%%%%%%%%%%%%%%%%%%%%%%%%%%
\bef
\begin{center}
\includegraphics[scale=0.4]{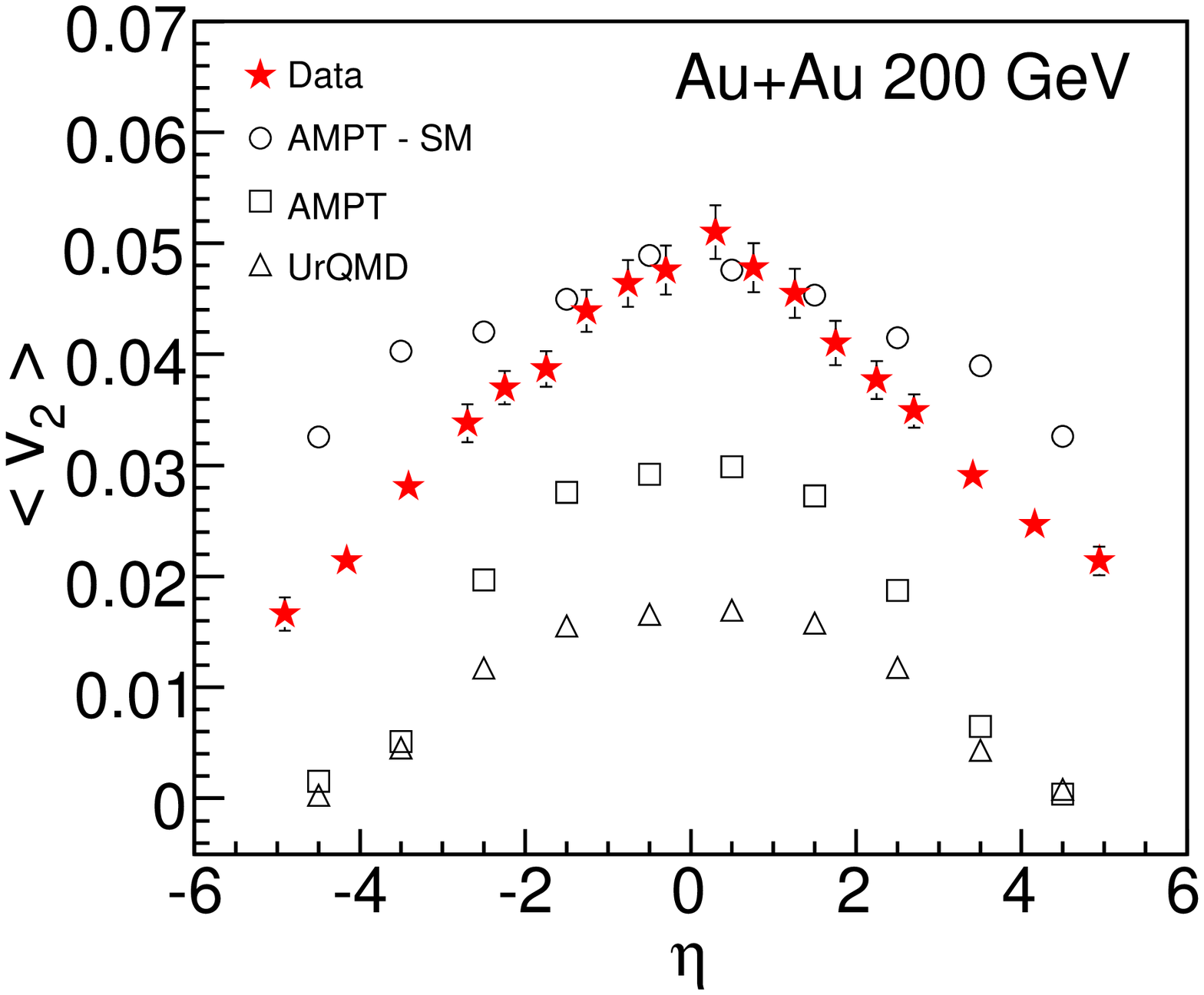}
\includegraphics[scale=0.4]{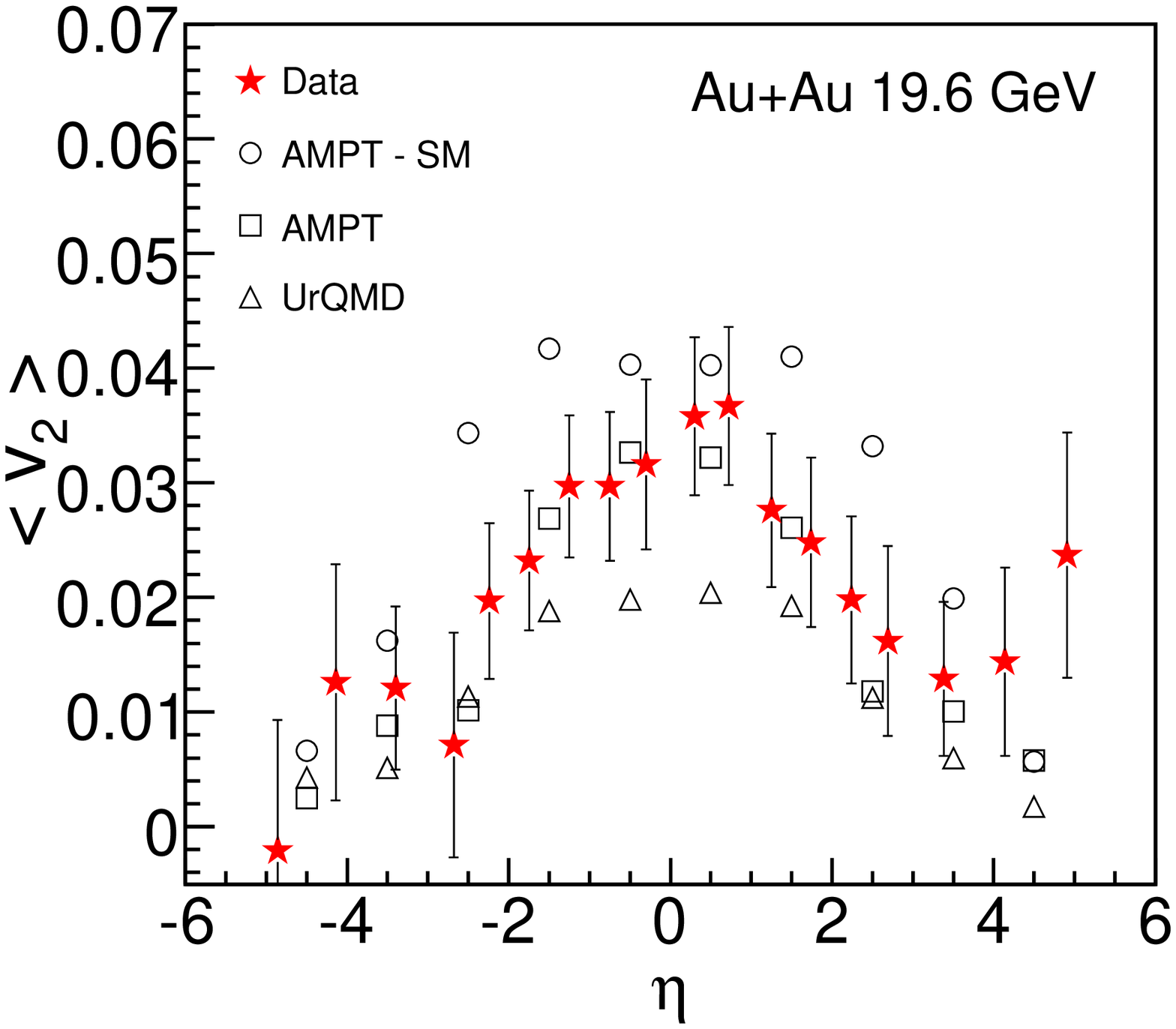}
\caption{(Color online) $\langle v_{2} \rangle $ of charged particles 
as a function of $\eta$ for Au+Au 0-40\% collisions 
at $\sqrt{s_{\mathrm {NN}}}$ = 19.6 and 200 GeV~\cite{phobosflow}.
The experimental results are compared to corresponding $v_{2}$ values from 
UrQMD and AMPT models. The errors on the data points are statistical.}
\label{fig4}
\end{center}
\eef
%%%%%%%%%% End of Fig.4 %%%%%%%%%%%%%%%%%%%%%%%%%%%%%
Figure~\ref{fig3} shows the comparison between experimental data on 
$v_{2}$  vs. $p_{\mathrm T}$ at midrapidity for Au+Au collisions and 
models at $\sqrt{s_{\mathrm {NN}}}$ = 200 and 9.2 GeV. The data for
200 GeV are for minimum bias 0-80\% collisions while those for 9.2
GeV are for minimum bias 0-60\% collisions. Reasonably good 
agreement between data and AMPT-SM is seen at 200 GeV, whereas the 
AMPT-SM model results are higher at 9.2 GeV. This is consistent with the 
results of $\langle v_{2} \rangle$ vs. $\sqrt{s_{\mathrm {NN}}}$ as 
shown in Fig.~\ref{fig2}.
Measurements exist for $\langle v_{2} \rangle$ vs. $p_{\mathrm T}$ at midrapidity 
for Au+Au collisions at $\sqrt{s_{\mathrm {NN}}}$ = 62.4 GeV, the comparison with
models is consistent with the trends as revealed by the energy dependence 
shown in Fig.~\ref{fig2}, hence not presented here.
The results from UrQMD and AMPT show similar trend of $v_{2}$ 
increasing with $p_{\mathrm T}$ but the values are lower compared to data at 200 GeV. 
These models however are in a reasonably good agreement with the data at
9.2 GeV. For UrQMD, this is consistent with the results shown in Fig.~\ref{fig2}.
The AMPT results on $\langle v_{2} \rangle$ at 9.2 GeV was found to be
higher than the data (Fig.~\ref{fig2}). The $\langle v_{2} \rangle$ folds the measured 
$v_{2}$  vs. $p_{\mathrm T}$ with the $p_{\mathrm T}$ distribution of
charged particles. We found the $\langle p_{\mathrm T} \rangle$ of 
charged particles from AMPT is larger than those from UrQMD, resulting in
the $\langle v_{2} \rangle$ being larger (as seen in Fig.~\ref{fig2}).

Figure~\ref{fig4} shows the comparison of the $\langle v_{2} \rangle$ vs. $\eta$
for charged particles (integrated over $p_{\mathrm T}$) from Au+Au
0-40\% collisions at $\sqrt{s_{\mathrm {NN}}}$ = 19.6 and 200 GeV~\cite{phobosflow}. 
The measurements
show a decrease in $\langle v_{2} \rangle$ as the $\mid\eta\mid$ increases.
The general trend of $\langle v_{2} \rangle$ vs. $\eta$ at 62.4 GeV is also similar
although the values lie intermediate to those from 19.6 and 200 GeV collisions.
All the models also qualitatively shows similar decreasing trend of $\langle v_{2} \rangle$
with increase in $\mid\eta\mid$, however they differ from the data in the 
widths of the $\langle v_{2} \rangle$ distribution in $\eta$ and the values 
of $\langle v_{2} \rangle$. 
AMPT-SM model agrees with the data at midrapidity for $\sqrt{s_{\mathrm {NN}}}$ = 200
GeV and is higher than the data at midrapidity for $\sqrt{s_{\mathrm {NN}}}$ = 19.6 GeV
(consistent with results in Fig.~\ref{fig2}).
However, for the higher rapidities the model fails quantitatively and has a larger
width of the $\langle v_{2} \rangle$  vs. $\eta$ distribution compared to data. The AMPT 
and UrQMD models fail to explain quantitatively the dependence of $\langle v_{2} \rangle$
on $\eta$ for $\sqrt{s_{\mathrm {NN}}}$ = 200 GeV. At $\sqrt{s_{\mathrm {NN}}}$ = 19.6 GeV
the AMPT  model is closer to the data values for all the $\eta$ range. 
The statistical errors on the data points are quite large. 

In summary, we have presented a compilation of the available data 
for elliptic flow parameter, $v_{2}$, of charged particles at RHIC  
as a function $\sqrt{s_{\mathrm {NN}}}$, $p_{\mathrm T}$ and $\eta$.
These results are compared to corresponding model calculations from
AMPT (default and string melting versions), UrQMD and HIJING.
The $\langle v_{2} \rangle$ values at midrapidity increases with
increase in $\sqrt{s_{\mathrm {NN}}}$. The $\langle v_{2} \rangle$ values 
from HIJING are consistent with zero at midrapidity for 
$\sqrt{s_{\mathrm {NN}}}$ = 19.6, 62.4, and 200 GeV. This suggests
that correlations among hadrons from minijets contribute negligibly to the
actual $\langle v_{2} \rangle$. The $\langle v_{2} \rangle$ results for 
the other models  studied do not show
a strong increase with $\sqrt{s_{\mathrm {NN}}}$.
The AMPT-SM model agrees quite well with the $\langle v_{2} \rangle$ data at 200 GeV
and gives higher values of $\langle v_{2} \rangle$ compared to data for the other
beam energies studied. The results from UrQMD match with the $\langle v_{2} \rangle$ 
data at 9.2 GeV and gives lower values of $\langle v_{2} \rangle$ compared to data for 
the other
beam energies studied. The calculated $\langle v_{2} \rangle$  values from AMPT  
model lie in between. Comparison with AMPT  shows the growing importance of 
initial and final state scattering to $\langle v_{2} \rangle$ with increase in 
$\sqrt{s_{\mathrm {NN}}}$. Considering the $\langle v_{2} \rangle$ values from 
UrQMD to be the contribution from hadronic matter, the higher values from the 
AMPT-SM model reflects additional contribution 
due to partonic interactions.The difference between UrQMD and data $\langle v_{2} \rangle$
values at midrapidity is about 60\% at 200 GeV. This together with agreement of the 
experimental measurements with AMPT-SM results at 200 GeV clearly indicates substantial 
contribution to the $\langle v_{2} \rangle$ measured at RHIC from partonic contributions. 
The agreement of measured $\langle v_{2} \rangle$ values with those from UrQMD 
at midrapidity for 9.2 GeV suggests that the experimental results can be understood within 
the frame work of a hadronic model.

We also presented a comparison of the measured $v_{2}$ vs. $p_{\mathrm T}$ and $\eta$
with the various models. The models qualitatively reproduce the trends of 
$v_{2}$ values increasing with increase in $p_{T}$ and decreasing with increase
in $\mid\eta\mid$. However, they lack in quantitative agreement at 200 GeV for the full 
$\eta$ range. Within the large statistical errors of the measured $v_{2}$ vs. $\eta$
at 19.6 GeV, we observed the AMPT model gives the best description of the
data. The RHIC beam energy scan program will make available the $v_{2}$ measurements
at energies of 7.7, 11.5, 18, 27 and 39 GeV. One can then further systematically study the
contribution of hadronic matter, initial state scattering and partonic matter 
to the $v_{2}$ measurements through comparisons with the above models discussed.

\noindent{\bf Acknowledgments}\\
We thank Zi-Wei Lin for useful discussions on AMPT model results.
Financial assistance from the Department of Atomic Energy, Government 
of India is gratefully acknowledged. PKN is grateful to the Board 
of Research on Nuclear Science and Department of Atomic Energy,
Government of India for financial support in the form of Dr. K.S. Krishnan
fellowship. LK is supported by DOE grant DE-FG02-89ER40531. \\

\end{document}